# Absence of magnetic proximity effects in magnetoresistive Pt/CoFe$_2$O$_4$ hybrid interfaces

M. Valvidares[1], N. Dix[2], M. Isasa[3], K. Ollefs[4], F. Wilhelm[4], A. Rogalev[4], F. Sánchez[2], E. Pellegrin[1], A. Bedoya-Pinto[3], P. Gargiani[1], L. E. Hueso[3,5], F. Casanova[3,5] and J. Fontcuberta[2]

[1] ALBA Synchrotron Light Facility, Carrer de la Llum 2-26, 08290 Cerdanyola del Vallès, Catalonia, Spain

[2] Institut de Ciència de Materials de Barcelona (ICMAB-CSIC), Campus UAB, 08193 Bellaterra, Catalonia, Spain.

[3] CIC nanoGUNE, 20018 Donostia-San Sebastian, Basque Country, Spain

[4] ESRF The European Synchrotron, CS40220, 71, Avenue des Martyrs, 38043 Grenoble Cedex9, France

[5] IKERBASQUE, Basque Foundation for Science, 48011 Bilbao, Basque Country, Spain.

*Corresponding authors: fontcuberta@icmab.cat; mvalvidares@cells.es*

Abstract

*Ultra-thin Pt films grown on insulating ferrimagnetic CoFe$_2$O$_4$ (111) epitaxial films display a magnetoresistance upon rotating the magnetization of the magnetic layer. We report here X-ray magnetic circular dichroism (XMCD) recorded at Pt-L$_{2,3}$ and Pt-M$_3$ edges. The results indicate that the Pt magnetic moment, if any, is below the detection limit (< 0.001 $\mu_B$/Pt), thus strongly favoring the view that the presence of CoFe$_2$O$_4$ does not induce the formation of magnetic moments in Pt. Therefore, the observed magnetoresistance cannot be attributed to some sort of proximity-induced magnetic moments at Pt ions and subsequent magnetic-field dependent scattering. It thus follows that either bulk (spin Hall and Inverse spin Hall Effects) or interface (Rashba) spin-orbit related effects dominate the observed magnetoresistance. Furthermore, comparison of bulk magnetization and XMCD data at (Fe,Co)-L$_{2,3}$ edges suggests the presence of some spin disorder in the CoFe$_2$O$_4$ layer which may be relevant for the observed anomalous non-saturating field-dependence of spin Hall magnetoresistance.*

## Introduction

The spin-orbit interaction) is at the heart of several magnetoresistance phenomena observed in metals. The anisotropic magnetoresistance (AMR) is one example and relies on the dependence of the charge carriers scattering on the direction of the local magnetization. Its angular dependence is well known and has been used in magnetic sensing for decades. In recent years however, it has been shown that spin-orbit interaction may have more subtle manifestations promoting, among other effects, pure spin currents and spin accumulation at the edges of non–magnetic metals (NM) in the presence of a charge flow (spin Hall effect,SHE) [1-3] or even unbalanced spin distributions at the symmetry-breaking metal surfaces and interfaces due to Rashba effect[4,5]. The presence of an external magnetic field[6] or a neighboring magnetic layer intimately coupled to the NM one[7-10] may modulate the spin accumulation and, via inverse spin Hall effect (ISHE), can be sources of magnetoresistance.

In fact, as shown by Nakayama et al.[7], if the metallic layer is placed in intimate contact with a magnetically ordered insulating layer, SHE and ISHE may combine to produce resistivity changes of the metallic layer depending on the orientation of the magnetic moments within the insulating ferromagnetic layer[7-9]. This magnetoresistance, named "spin Hall magnetoresistance" (SMR)[10], is receiving much attention in the quest for spin-only devices. Magnetoresistance in paramagnetic metallic layers grown onto ferromagnetic insulators has been identified in Pt/$Y_3Fe_5O_{12}$ (YIG) bilayers, and several other bilayer systems including metals such as Ta[8, 11, 12] or Pd[13] and magnetic insulating thin films such as $Fe_3O_4$[14, 15], $NiFe_2O_4$[14], $CoFe_2O_4$[16, 17], $SrMnO_3$[18] or $CoCr_2O_4$[19].

However, assigning any measured magnetoresistance in the metallic layer, say Pt, to SMR or interface Rashba field is challenging, as the Pt may become spin polarized by proximity effect, prompting a radically different picture[20-22]. Separation of these two physical origins relies on the assessment of whether magnetic moments have been induced or not in the metallic layer by its magnetic neighbor. As a matter of fact, it is well known that not only in metal-metal interfaces (i.e. Ni/Pt)[23] but also in metal-insulator interfaces (i.e. Co/$LaFeO_3$)[24] magnetic moments can be induced across the interface into the non-magnetic phase.

In the context of magnetoresistance of the most studied hybrid Pt/YIG bilayers, the eventual presence of induced moments in the Pt layer has not yet received an unambiguous answer. Indeed, while all transport experiments agree on the presence of an induced magnetoresistance in the Pt layer, arguments have been put forward claiming for proximity-induced magnetic moments in Pt as the reason for the observed magnetoresistance[25], although others denied this conclusion and supported a spin-Hall-related origin[26]. Element sensitive X-ray magnetic circular dichroism (XMCD) has been used to determine the magnetic moment of Pt in Pt/YIG bilayers. Geprägs et al. reported XMCD data at Pt-$L_{2,3}$ edges[26] of Pt(3 nm)/(111)YIG(62 nm) bilayers where the Pt and YIG layers were grown by electron beam-evaporation and pulsed laser deposition (PLD) respectively on $Y_3Al_5O_{12}$ (YAG). Upon comparison with the XMCD signals obtained from Pt/Fe bilayers, they concluded that the magnetization of the Pt layer, if any, should be limited to about ≈ (0.003± 0.001) $\mu_B$/Pt integrated over the complete Pt thickness, much smaller than that observed in Pt/Fe interfaces[26]. In contrast, Lu et al.[25] reported ≈ 0.054 $\mu_B$/Pt in Pt(1.5 nm) films grown on 18 $\mu$m thick (111)YIG layers prepared by liquid phase epitaxy on $Gd_3Ga_5O_{12}$ (GGG) substrates.

Recently, it has been found that bilayers formed by Pt and insulating spinel ferrites ($CoFe_2O_4$, $NiFe_2O_4$, etc.) display an angular dependent magnetoresistance compatible with SMR and interface-Rashba mechanisms, both having the same angular variations.[14, 16, 4] Here, to settle if the observed effects are a signature of spin accumulation at the interface or of magnetic moment formation by proximity effect, we report on XMCD measurements at the Pt-$L_{2,3}$ and Pt-$M_3$ edges in Pt/$CoFe_2O_4$ (Pt/CFO) heterostructures complemented with bulk magnetometry and magnetotransport measurements. We show that, whereas a clear magnetoresistance is observed in the Pt layer with an angular dependence fully consistent with predictions for SMR, a negligible magnetic moment at the Pt atoms (< 0.001 $\mu_B$/Pt) is derived from the XMCD data. Therefore, we conclude that the magnetic response of the Pt layer grown on the ferrimagnetic CFO film does not originate from proximity-induced magnetism at Pt ions but from bulk or interface spin-orbit effects. XMCD measurements were also performed at (Fe,Co)-$L_{2,3}$ edges, thus providing a unique complementary information on the magnetic properties of the heterostructures. Results indicate the presence of non-saturated ferromagnetic regions in the CFO layer accounting for the observed non-saturated high-field behavior of the SMR.

**Experimental details**

CFO films were epitaxially grown on (111) SrTiO$_3$ (STO) substrates by PLD using a KrF laser with a fluence around 1.5 J/cm$^2$ and a repetition rate of 5 Hz at a temperature of 550 °C and oxygen pressure $P_{O2}$ = 0.1 mbar[27]. Pt layers (7 nm) were deposited on the CFO layers by dc sputtering at 400° C. Three different samples were prepared: a) STO//CFO (40 nm)/Pt, b) STO//CFO (28 nm)/Pt and c) STO//CFO (28 nm)/Pt/CFO (28 nm). All layers were grown in a UHV set-up that allows sample transfer from the PLD to the sputtering chambers preserving UHV conditions at all times. The STO//CFO(40 nm)/Pt sample was patterned into Hall bars (width W=100 μm and length L = 800 μm), for measuring magnetoresistance[16]. Resistivity of Pt is typically ≈20 μΩcm. The STO//CFO (28 nm)/Pt bilayer, having an exposed Pt surface, was used for X-ray absorption (XAS) experiments of Pt edges, whereas the trilayer, having an exposed CFO surface, was used for XAS at Fe and Co edges.

Magnetization measurements were done using a SQUID magnetometer. Magnetotransport measurements were performed at 100 K with external magnetic field (H) ranging from -90 kOe to +90 kOe applied at different angles. XAS and XMCD measurements were performed at the soft X-ray Pt-$M_3$ and (Fe, Co)-$L_{2,3}$ edges at the BOREAS BL29 beamline of the ALBA Synchrotron Light Facility, using total electron yield (TEY) detection. Ultimate probing of magnetic moments at the Pt electrodes was achieved by measuring the hard X-ray Pt-$L_{2,3}$ edges at the ID12 beamline of the European Synchrotron Radiation Facility (ESRF), which offers extremely high sensitivity at high photon energies. In the latter measurements, total fluorescence yield detection mode was used for the collection of XAS spectra at the Pt-$L_3$ (11567 eV) and Pt-$L_2$ edges (13271 eV).

**Results and Discussion**

As an illustrative indication of the samples quality, we include in **Figure 1** the X-ray diffraction θ-2θ pattern of the most complex heterostructure studied: STO//CFO(28 nm)/Pt/CFO(28 nm). The (*hhh*) reflections of CFO are well appreciated and indicate a c-axis length of 8.39 Å. The Laue fringes of the thin Pt layer are also well visible. The splitting of the Laue fringes gives a Pt thickness of about 7.3 nm, in close agreement with the value expected from calibrated Pt growth rate.

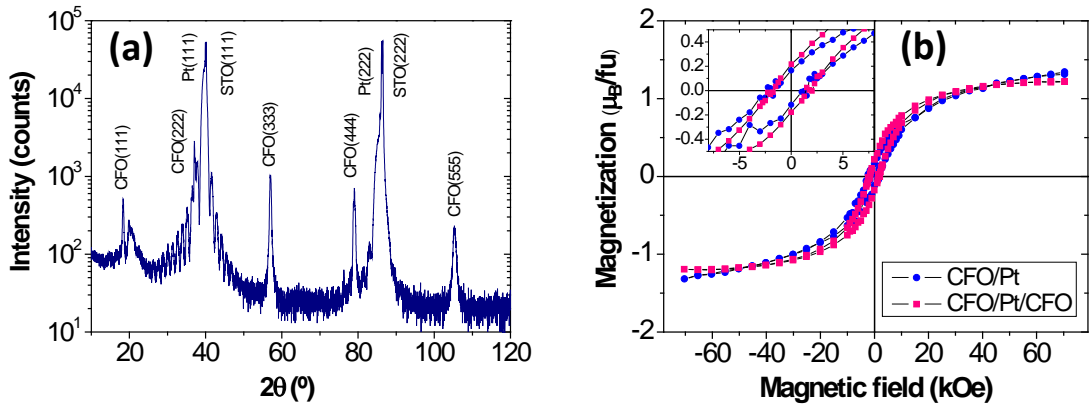

**Fig. 1.** *a)* X-ray diffraction θ-2θ pattern of the STO//CFO(28 nm)/Pt/CFO(28 nm) sample; *b)* room-temperature magnetization loops for the STO//CFO(28 nm)/Pt/CFO(28 nm) and STO//CFO(28 nm)/Pt samples. Inset in *b)* is a zoom of the low-field region (units as in main panel) of the magnetization loops.

We show in Fig. 1b the room-temperature magnetization M(H) loops of the STO//CFO(28 nm)/Pt and STO//CFO(28 nm)/Pt/CFO(28 nm) heterostructures measured with the magnetic field applied in the film plane. The diamagnetic STO signal is roughly eliminated by subtracting a linear contribution extrapolated from the high-field region of the raw data. Results reveal features commonly found in spinel oxide films, namely a small remanence and a reduced saturation magnetization. Indeed, the magnetization at the highest field is smaller than that corresponding to an ideal (fully inverse) cationic distribution, assuming a spin-only magnetic moment of high-spin $Co^{2+}$(S = 3/2) ions at the octahedral B-sites in the CFO layer ($3\mu_B$/f. u. ≈ 376 emu/cm$^3$). A partial inversion $(Fe_{1-x}Co_x)_T [Fe_{1+x}Co_{1-x}]_O O_4$ (x > 0) (here the sub-scripts "O" and "T" indicate octahedral and tetrahedral sites, respectively) would lead to saturation magnetization values even larger. The observation of a reduced magnetization is usually attributed to: i) the presence of antiphase boundaries formed during thin film growth that introduce hard-to-saturate antiferromagnetic regions in the film[28-30] and ii) the presence of surface anisotropy [31], both effects contributing to a slow approach to saturation. We note in passing that, although the measurements reported here have been done at 300 and 100 K and the magnetization should further increase at the lowest temperature, the fact that the Curie temperature of CFO films has been reported to be as high as 840 K[32] and values ranging from 520 K to 683 K are quoted for bulk CFO by J. Smit and H.P. Winjn[33] and S. Chikazumi[34], respectively, indicate that the expected thermally-induced increase of magnetization down to 0 K cannot be larger than about 10%, which is insufficient to explain the observed reduction and consequently other mechanisms cooperate and rule the observed suppression of magnetization.

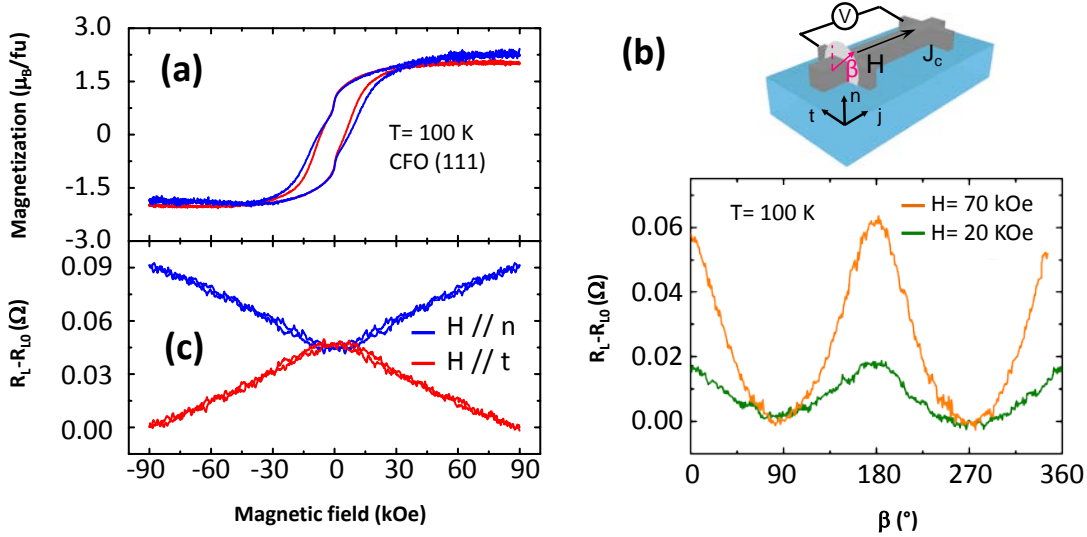

**Figure 2.** *(a) Magnetization loops for the STO//CFO(40 nm)/Pt sample with **H** applied along **n** (blue curve) and **t** (red curve) as sketched in (b). (b) top: Sketch of the experimental configuration; bottom: angle-dependent longitudinal magnetoresistance at 20 kOe and 70 kOe, when rotating **H** in a plane perpendicular to the current as shown in the sketch. $R_L$ is the measured resistance. (c) Field-dependent $R_L$ with **H**, applied along **t** (red) and **n** (blue). In b) and c) $R_{L0}$ are (subtracted) resistance backgrounds.*

In **Figure 2a** we show M(H) loops of the STO//CFO(40 nm)/Pt sample recorded at 100 K, with **H// t** and **H//n** (**t** and **n** are unit vectors transverse to the current path within the film plane and perpendicular to it, respectively). It can be appreciated that the coercive fields along these two perpendicular directions are almost coincident; this observation indicates that shape anisotropy is not prevalent in the film and it is consistent with the fact that in CFO films on STO the easy axis is along the [100] direction [35] thus implying identical (assuming a cubic CFO unit cell) projections along **t** and **n** directions plane. In **Figure 2b** we show the longitudinal magnetoresistance $R_L(\beta)$ measured at 100 K, when the field is rotated in a plane perpendicular to the measuring current. $\beta$ is the angle between the applied field **H** (20 kOe and 70 kOe) and the normal to the film. The data display the $\cos^2(\beta)$ dependence expected for Rashba-induced magnetoresistance[4] and for SMR[16]. We stress that in this measuring configuration, the AMR contribution should vanish and SMR should saturate when the magnetization of the layer saturates. However, the SMR does not. This is confirmed in Fig. 2c, where $R_L(H)$ curves recorded at **H//t** and **H//n** do not saturate up to 90 kOe whereas the M(H) (see Fig. 2a) displays only a small differential susceptibility in the 50-90 kOe range. At this point, it is important to note that, as mentioned above when referring to the magnetization data of Fig. 2a, the substrate contribution is eliminated by subtracting the linear high-field magnetization data. This commonly used procedure unavoidably entangles any high-field susceptibility of the film with the large diamagnetic contribution of the substrate, thus hiding any intrinsic high-field susceptibility of the film and challenging an accurate determination of film magnetization. The SMR, being insensitive to substrate contribution, shows in the crudest way a large high-field slope which originates from the film magnetization. To get information on the genuine magnetic properties of the CFO layers, we have performed XMCD measurements at the (Fe,Co)-$L_{2,3}$ core levels on STO//CFO(28 nm)/Pt/CFO(28 nm) sample. For this purpose we have used two different photon energy beam lines. At the soft X-ray energy beam (at ALBA synchrotron) allows inspection of L-Fe,Co edges and with a reduced sensitivity the M-Pt edge, whereas a higher energy beam at ESRF allows access more sensitive L-Pt edge. Results are described consecutively described in the following.

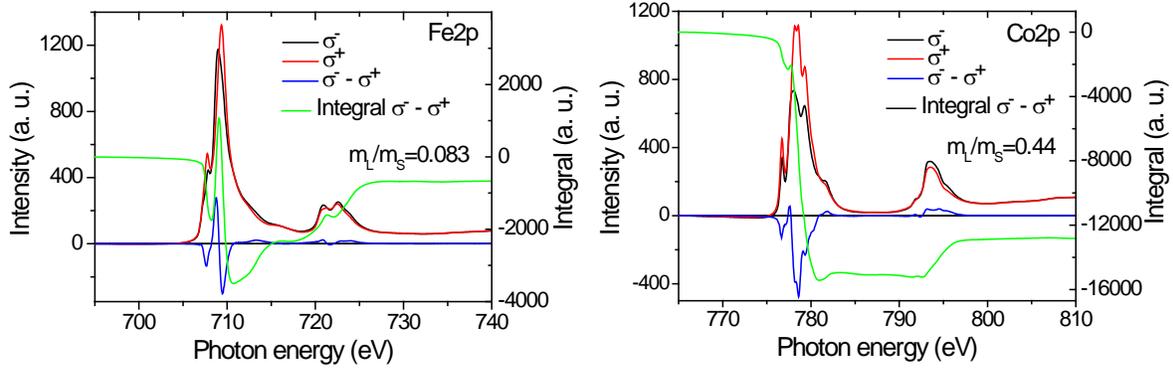

**Figure 3.** *(a) Fe-$L_{2,3}$ and (b) Co-$L_{2,3}$ 2p XMCD spectra STO//CFO(28 nm)/Pt/CFO(28 nm) sample (295 K, 60 kOe, normal incidence). The black, red and blue lines (left axis) show the absorption for $\sigma^-$ and $\sigma^+$ photon helicities and the dichroic ($\sigma^-$ - $\sigma^+$) spectra, respectively. The green line shows the integrated difference spectrum (right axis).*

In **Figures 3a and 3b,** we show the (Fe,Co)-$L_{2,3}$ XMCD spectra of the STO//CFO(28 nm)/Pt/CFO (28 nm) sample collected at room temperature under a field of 60 kOe at normal beam incidence (H parallel to the beam). The corresponding magnetization loops are shown in **Fig. 1b.** The integrated areas under the corresponding dichroic signals (right axes) allow extracting the spin and angular parts of the magnetic moment and the corresponding $m_L/m_S$ ratio. It turns out that $m_S(Fe^{3+})$ = 0.31 $\mu_B$, $m_L(Fe^{3+})$ = 0.026 $\mu_B$, and $m_S(Co^{2+})$ = 1.19 $\mu_B$, $m_L(Co^{2+})$ = 0.525 $\mu_B$, where these values correspond to the net *total* moment averaged over the different cation species at the tetrahedral and octahedral sites of $(Fe_{1-x}Co_x)_T [Fe_{1+x}Co_{1-x}]_O O_4$. In this context, one has to note the well-known limitations pertinent to the spin sum rule for the determination of the effective spin moment[36] that can be taken into account by the introduction of correction factors, which have been reported to be on the order of about 8% for the late transition metals such as $Co^{2+}(3d^7)$, and about 31% for $Fe^{3+}(3d^5)$ systems. Applying these corrections to the above averaged spin moments, we obtain $m_S(Fe^{3+})$ = 0.44 $\mu_B$ and $m_S(Co^{2+})$ = 1.3 $\mu_B$. From these corrected spin values and the corresponding orbital magnetic moments, we estimate $M_S \approx 2.76$ $\mu_B$/f.u. CFO. We note that this value is larger by about a factor of two than that derived from the SQUID data (Fig. 1b), illustrating how critical is the subtraction of the substrate contribution from the measured magnetic moment in SQUID measurements and/or differences in averaged (SQUID) and surface (XMCD) film magnetization. However, it is still smaller than that expected value for a fully inverse, spin-only, CFO spinel ($M_S$ = 3 $\mu_B$/f. u.) if all atomic magnetic moments were aligned along the magnetic field axis. Last, but not least, we would like to note that equivalent XMCD measurements on a $CoFe_2O_4$ single crystal did yield a total magnetic moment of 3.23 $\mu_B$/f.u.[37].

Of major interest here are the XMCD data recorded at the Pt-$M_3$ and Pt-$L_{2,3}$ absorption edges, which probe the magnetism of Pt atoms with element specificity. **Figures 4a** and **4b** show respectively the room-temperature soft x-ray XAS and XMCD signals measured at Pt-$M_3$ edge of the STO//CFO(28 nm)/Pt bilayer using TEY at normal incidence and for a 70% circular polarization.

These measurements indicate that there is no appreciable magnetic dichroism under an applied magnetic field of 60 kOe. For comparison, we measured a SiN membrane supported (Co(0.5nm)/Pt(0.5nm))$_3$ multilayer which clearly evidences a Pt magnetic moment by an XMCD with magnitude of about 5% percent of the Pt-$M_3$ white line intensity. The differences between the main edge peak structure apparent between Pt XANES in Figs. 4a and 4b arises mainly due to the different absorption backgrounds present in the Pt/CFO//STO structure as compared to the much simpler Pt/Co/SiN-membrane sample. The latter membrane sample substrate itself has weak background and consists of Pt, Co ultra-thin layers only, whereas the former spin hall structure includes much more elements and a larger film thickness, thus resulting into a much larger non-resonant background contributions that superpose the Pt M3 edge.[38] In the following, we provide some considerations that allow estimating an upper bound value for the vanishing Pt-$M_3$ XMCD and its corresponding magnetic moment.

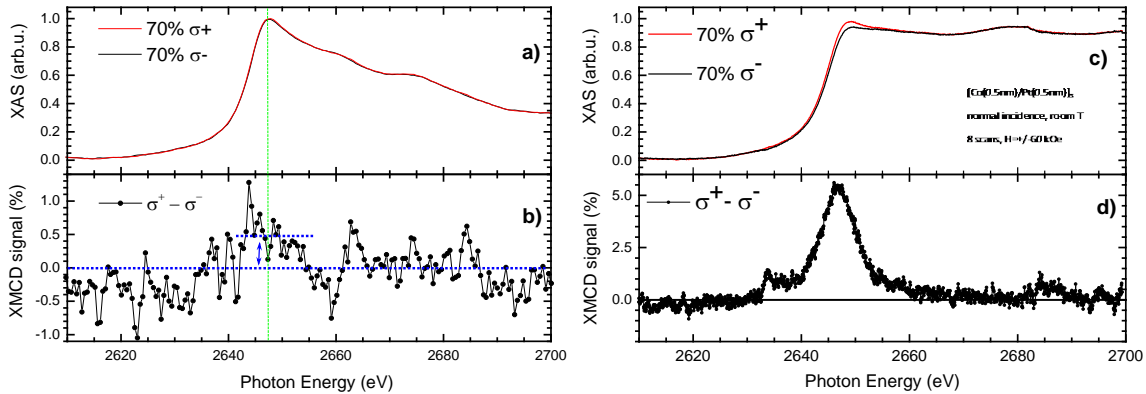

**Figure 4.** *(a) X-ray absorption spectra for photon ($\sigma^+$ and $\sigma^-$) helicities and (b) ($\sigma^+ - \sigma^-$) dichroism at the Pt-$M_3$ edge for the STO//CFO(28 nm)/Pt sample. XAS has been normalized to 1.0 at the white line peak; (b) XMCD signal scaled as percentage of the white line intensity at about 2649 eV. The red arrow indicates the average (≈ 0.4(7)%) XMCD value at the white line photon energy region (short black dotted line) with respect to the zero XMCD baseline (long black dotted line).The corresponding XAS and XMCD on a reference (Co/Pt)$_3$ multilayer in (c) and (d) illustrates the measurement sensitivity for probing Pt magnetic moment at Pt-$M_3$ edge.*

To establish a suitable calibration for XMCD at the rarely used soft x-ray photon energy of the Pt-$M_3$ edge, the (Co/Pt)$_3$ reference multilayer serve only as a rough estimate because Co/Pt interface magnetic moments can depend strongly on the interface quality and degree of alloying. If we assume a value for the induced Pt moment of 0.3 $\mu_B$/Pt which is typical of multilayers[39], that would calibrate the XMCD at Pt-$M_3$ edge at roughly 0.3 $\mu_B$/Pt ×(1/5%), i.e. 0.06 $\mu_B$/Pt per 1% XMCD at Pt $M_3$ edge. An alternative calibration could be based on a better defined reference such as a bulk alloy, for what we note that in CoPt$_3$ alloys Grange et al.[39] reported a XMCD ≈ 10% at the Pt-$L_3$ edge for a magnetic moment of ≈ 0.3 $\mu_B$/Pt. Therefore, considering the XMCD at $L_3$ and $M_3$ edges are in an approximate ratio ≈5/1 [40], we can estimate that a Co-Pt sample, having a magnetization of ≈ 0.3 $\mu_B$/Pt would display a XMCD ≈ 2 % at the Pt-$M_3$ edge. Taking into account that our Pt-$M_3$ measurements were performed using 70% circularly polarized light, the expected XMCD at M-edge signal for this given Pt magnetic moment would be about 1.4%, yielding an approximate calibration of 0.21 $\mu_B$/Pt per 1% XMCD at the Pt-$M_3$ edge.

In our measurements shown in Fig. 4b, if we assume XMCD is essentially zero over the whole energy range, the noise (standard deviation of XMCD data) amounts to 0.3%. If there would be any XMCD at the Pt-$M_3$ edge, it would be expected in a region interval around the Pt-$M_3$ peak maxima (as in the case on Fig. 4d): considering the range 2629 to 2641 eV, the mean value for XMCD is ~ 0.5% with an standard deviation of~ 0.3%. Therefore, one could consider this 0.4(7)% mean value as a statistical representative value for XMCD in this sample at the Pt-$M_3$ edge. The Pt-$M_3$ region and its statistical XMCD mean value are indicated with a short black dotted line in Fig. 4. According to the approximate calibrations established above, this would set an upper bound for thickness-averaged magnetic moment of the Pt layer in the Pt/CoFe$_2$O$_4$ bilayer between 0.5% × 0.06 $\mu_B$-Pt/% = 0.03 $\mu_B$/Pt and 0.5% × 0.21 $\mu_B$-Pt/% = 0.10$\mu_B$/Pt. The resulting 0.03-0.10 $\mu_B$ range of values for the Pt layer-averaged magnetic moment supports that in our Pt/CoFe$_2$O$_4$ bilayer the proximity effects are very weak or eventually absent. This discussion also evidences that in spite of the rather good zero measurements at the Pt-$M_{2,3}$ edges, further measurements with increased sensitivity are needed to push the upper bound of Pt magnetic averaged moment to even lower values.

As mentioned, we performed complementary measurements of the STO//CFO(28 nm)/Pt sample at the Pt-$L_{3,2}$ edges at the ESRF ID12 beamline, which allowed to benefit from the larger Pt-$L_{3,2}$ edges XMCD sensitivity combined with the use of a very efficient lock-in based detection scheme. In **Figure 5** (left axis) we show the near-edge X-ray absorption (XANES) spectra around the Pt-$L_{2,3}$ edges collected at grazing incidence (5°), measured with fluorescence yield and under H = 9 kOe at room temperature. We note that the XANES spectrum is typical of metallic Pt[41], in particular evidencing clear lineshape wiggles at energies (11587 eV, 11300 eV) right above $L_3$, $L_2$ edges which are well known to be characteristic of metallic Pt[42]. More quantitatively, following Gepraegs et al*[42]*, it is to be noted that the Pt $L_3$ whiteline intensity in our sample is of about 1.24, a value closely matching that found for metallic Pt (1.25) and definitely much smaller that of the corresponding Pt edge in PtO$_x$[43] This denies oxidation of Pt due to the relatively high deposition temperature (400°C) or implantation due to the sputtering process, and is indicative of a clean and sharp high-quality Pt interface.

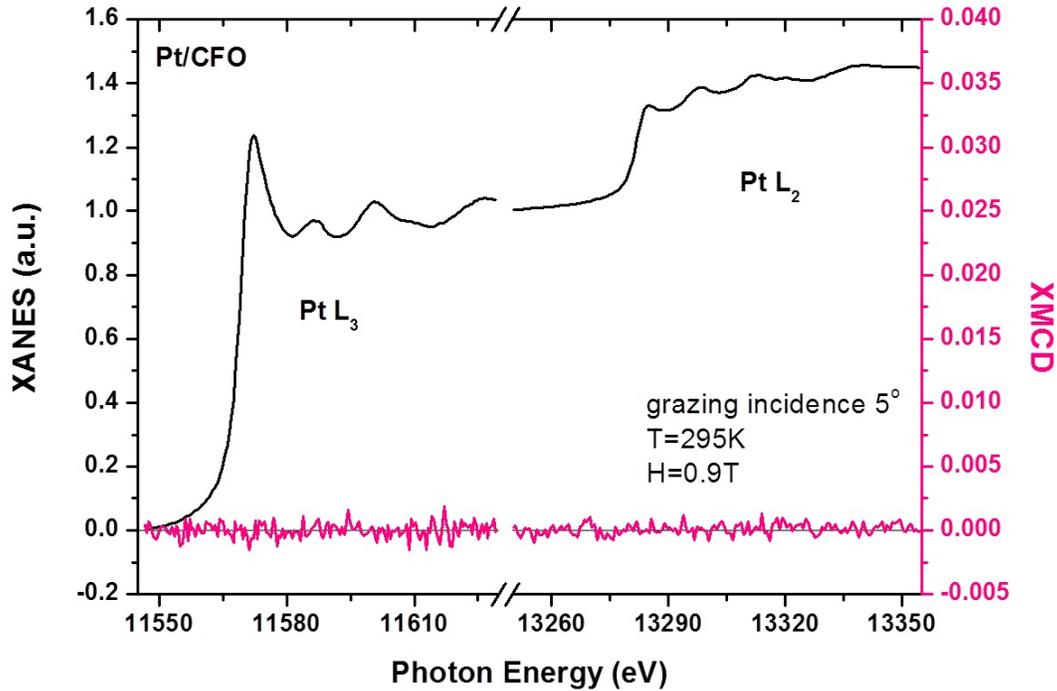

**Figure 5.** X-ray absorption spectra (black line, left axis) measured across the Pt-$L_{3,2}$ atomic absorption edges of the STO//CFO(28 nm)/Pt sample. Right axis: X-ray Magnetic Circular Dichroism signal.

In Figure 5 (right axis) we show the corresponding XMCD. No XMCD signal is apparent, remaining within the noise level. This implies that any magnetic moment at the Pt atoms should be ≤ 0.001 $\mu_B$/Pt averaged over the Pt layer thickness. It is worth to recall that the used experimental arrangement at ESRF beamline ID12 has a well proven record of ultimate sensitivity to Pt moments: it has been used to measure the XMCD signal at the Pt-$L_3$ edge in Pt(3 nm)/Fe(10 nm) bilayers, indicating an induced Pt magnetic moment of 0.03 $\mu_B$/Pt averaged over the complete Pt film thickness[26]. It follows that in our STO//CFO(28 nm)/Pt film using the Pt-$L_3$ edge we can set an upper value for the average Pt magnetic moment (0.001 $\mu_B$/Pt). This value is more than one order of magnitude smaller than that found for the Pt/Fe interface (≤ 0.03 $\mu_B$/Pt) and about 50 times smaller than the Pt magnetic moment previously reported in Pt/YIG bilayers (m(Pt) ≈ 0.05 $\mu_B$/Pt)[25]. We notice that, in spite of the layer averaging characteristics of fluorescence, which has been taken into account in the analysis, the sensitivity of the measurement is so high that still sets a very negligible value (upper bond) for any Pt magnetic moment in any point of the layer. We also stress that for close to optimum magnetoresistance structures, Pt layers have thickness in the range of 5 to 10 nm, which give fluorescence a similar or higher sensitivity to the very Pt/CFO interface than a total electron yield approach.

Before concluding, we should add that XMCD experiments are only sensitive to magnetic moments that project along the beam direction, which is also the direction of the applied magnetic field, and experiments at ESRF ID12 beamline have been performed at grazing incidence. In presence of surface anisotropy affecting Pt moments, the present results could not be more than an upper bound to any possible induced magnetic moment. However, at the magnetic field used in the ESRF XMCD experiments (9 kOe), according to the magnetic data in Fig. 1b, the CFO magnetization is already at the 50% of its highest value, and thus a correspondingly adapted upper limit could be a factor of two larger, i.e., m(Pt) ≤ 0.002 $\mu_B$/Pt.

**Summary and Conclusions**

In summary, XMCD measurements at the Fe- and Co-$L_{2,3}$ edges of in STO//CFO(28 nm)/Pt/CFO(28 nm) heterostructures give clear evidence of a reduced magnetization, thus supporting the view that antiphase boundaries limit the CFO magnetization both in the film bulk and at its surface. As a natural consequence, the film magnetization approaches saturation in a very slow manner and this observation provides a simple explanation for the observation that the magnetoresistance increases with field (Fig. 2c) at field values where the bulk magnetization loops appear already rather saturated. As SMR is a genuine interface effect limited by the spin mixing conductance across interfaces, it is extremely sensitive to interface magnetism. Importantly the XMCD measurements at the Pt-$L_{2,3}$ in STO//CFO(28 nm)/Pt set an upper bound for Pt magnetic moment of ≤ 0.002 $\mu_B$/Pt. This observation would indicate that magnetic proximity effects in this interface are negligible, thus supporting the view that the observed magnetoresistance of the Pt layer is due to either SMR or Rashba field; as mentioned the angular dependence of the magnetoresistance alone does not permit to discriminate among these different scenarios. In any event, the results reported here hold in a set of high quality, PLD-grown CFO layers with *in-situ* (UHV) sputtered Pt overlayers; it should not be a surprise that differences on interface structure and quality, density of antiphase boundaries or other morphological and structural properties may impact SMR and proximity effects, eventually yielding different results as might be the case for YIG systems[44-47]. After completion of this manuscript, T. Kuschel et al.[48] reported X-ray resonant magnetic reflectivity data on Pt/NiFe$_2$O$_4$ and concluded that the magnetic moment at Pt should be below 0.02 $\mu_B$/Pt. Our present data set a lower upper bound for Pt moment about one order of magnitude lower, which is agreement also with a recent report[17].

In this manuscript, we have focused on the spin magnetoresistance originating from current-induced pure spin currents. Spin currents can also be originated, among others, by thermal gradients giving rise to spin Seebeck effects[49]. In this context, hybrid Pt/CoFe$_2$O$_4$ bilayers have also been recently explored and shown that thermally generated spin currents can also diffuse across Pt/CoFe$_2$O$_4$ interfaces [50, 51]. It thus may not be a surprise that similar effects to those reported here could be detected, namely a nonmagnetic saturation of the thermally generated currents, and thus of the corresponding Seebeck voltage in the Pt probing contacts, while the film magnetization loops appear to be saturated.

**Acknolwedgements**

This work is supported by the European Union under the NMP project (263104-HINTS) and the European Research Council (257654-SPINTROS). The financial support from the Spanish Ministry of Economy and Competitiveness, through the "Severo Ochoa" Program for Centers of Excellence in R&D (SEV-2015-0496) and the projects (MAT2012-37638, MAT2014-56063-C2-1-R, FIS2013-45469-C4-3-R, MAT2014-59315-R and MAT2015-65159-R) and by the Catalan Government (2014 SGR 734) are acknowledged. M. I. acknowledges the Basque Government for a PhD fellowship (BFI-2011-106). Beam time access at ESRF and ALBA facilities, at beamlines ID12 and BL29 respectively, is also acknowledged.